\documentclass[reprint,aps,prl]{revtex4-1}

\usepackage{graphicx,amsmath}

\usepackage{amsmath}
\usepackage{epsf}
\usepackage{graphicx}

\usepackage{dcolumn}
\usepackage{bm}

\usepackage{amssymb}
\usepackage{array}
\usepackage{varioref}
\usepackage{wrapfig}
\usepackage{etoolbox}
\usepackage{color}

\providecommand{\nn}{\nonumber}

\providecommand{\bv}[1]{\bm{\mathrm{#1}}}

\providecommand{\w}{\omega}
\providecommand{\W}{\Omega}
\providecommand{\q}{\bv{q}}

\renewcommand{\k}{\bv{k}}

\providecommand{\ef}{\varepsilon_F}
\providecommand{\vf}{v_F}

\providecommand{\kf}{k_F}

\providecommand{\kf}{k_F}

\renewcommand{\q}{\bv{q}}

\providecommand{\gb}{\bar{g}}

\providecommand{\Sg}{\Sigma}

\providecommand{\sgn}{\mbox{sgn}}

\providecommand{\sgn}{\mbox{sgn}}
\newcommand{\beq} {\begin{equation}}
\newcommand{\eeq} {\end{equation}}
\newcommand{\bea} {\begin{eqnarray}}
\newcommand{\eea} {\end{eqnarray}}
\renewcommand{\ef}{E_F}
\begin{document}

\title{
Superconductivity near a nematic quantum critical point -- the interplay between hot and lukewarm regions }

\author{Avraham Klein}
\author{Andrey Chubukov}
\affiliation{School of Physics and Astronomy, University of Minnesota, Minneapolis. MN 55455}

\begin{abstract}
  We present a strong coupling dynamical theory of the superconducting transition in a metal near a QCP towards $Q=0$ nematic order. We use a fermion-boson model, in which we treat the ratio of effective boson-fermion coupling and the Fermi energy as a small parameter $\lambda$. We solve, both analytically and numerically, the linearized Eliashberg equation. Our solution takes into account both strong fluctuations at small momentum transfers $\sim \lambda k_F$
  and weaker fluctuations at large momentum transfers.
  The strong fluctuations determine $T_c$,
  which is of order $\lambda^2\ef$ for both $s-$ and $d-$ wave pairing.
  The
  weaker fluctuations determine the
  angular
  structure of the
  superconducting order parameter
  $F(\theta_k)$ along the Fermi surface, separating between hot and lukewarm regions.
  In the hot regions $F(\theta_k)$ is largest and approximately constant. Beyond the hot region, whose width is $\theta_h \sim \lambda^{1/3}$,
  $F(\theta_k)$
  drops
  by a factor $\lambda^{4/3}$.
  The $s-$ and $d-$ wave states are not degenerate
  but the relative difference $(T^s_c-T^d_c)/T^s_c \sim \lambda^2$ is small.

\end{abstract}
\maketitle

\paragraph*{{\bf Introduction}}
\label{sec:introduction}

Superconductivity (SC)  mediated by fluctuations arising from proximity to an electronic quantum-critical point (QCP) has attracted tremendous interest in the ``high $T_c$'' era.
Much of the motivation comes from the known proximity of the Cu- and Fe- based superconductors to antiferromagnetism~\cite{Loehneysen2007,Monthoux2007,Abanov2003,Scalapino2012,Sachdev2012,Cyr-Choiniere2018}
but more recent discoveries of charge-density-wave order in the cuprates and of nematic order in both Cu-and Fe-based materials\cite{Taillefer2010,Shibauchi2014,Coldea2017}
have
led to studies of SC mediated by critical charge fluctuations~\cite{Fernandes2014,Wang2015,Wang2014}.
Theoretical studies of SC near a QCP  show that it is a strong coupling phenomenon, arising from the divergent fluctuations \cite{Abanov2001,Wang2016,Raghu2015}.
These
 fluctuations also induce large electronic self-energies, which in the absence of SC would account for a non Fermi liquid (NFL) behavior below some characteristic frequency $\omega_0$ \cite{Altshuler1994,Abanov2001,Abanov2003,Metlitski2010a,Metlitski2010,Metlitski2010b,Lederer2017,Lee2018}.
In some systems SC emerges at
$T_c \gg \omega_0$ and masks
 the
NFL behavior~\cite{Metlitski2015,Raghu2015}, in other systems $T_c$ is smaller than $\omega_0$, at least numerically. In the latter case SC emerges out of a NFL.

A subset of  theories of SC in a quantum-critical regime are those dealing with transitions at vanishing momentum transfer $Q = 0$ \cite{Bonesteel1996,Altshuler1994,Nayak1994,Nayak1994a,Fradkin2010,Metlitski2010a,Rech2006,Lee2009,Lee2018,Raghu2015,Metlitski2015}.
They are typically associated with a deformation of the Fermi surface (FS) in some angular momentum channel, e.g. $l=2$ for the nematic transition of the type observed in Fe- and Cu-based SCs. A theory of pairing mediated by soft fluctuations of $d-$wave nematic order parameter must account both for the strong coupling
 physics that occurs locally on the Fermi surface (FS), and for the momentum  anisotropy caused by a $d-$wave form-factor, which occurs on the large momentum scale  of the Fermi wavevector $\kf$.

This paper deals with SC at the nematic QCP.
The $\cos{2\theta}$ form of the $d-$ wave form-factor
splits the FS into four `hot' regions where $\theta \approx n \pi/2$, $n= 0,1,2,3$, where interactions are strong, and four 
`lukewarm'
regions where $\theta \approx (n+1/2)\pi/2$, where the pairing interaction is much weaker~\cite{Lederer2015,Schattner2016}.  Previous studies of this problem have focused either on the pairing away from a QCP within a Fermi liquid framework~ \cite{Lederer2015}, or on local strong-coupling properties in the hot regions~\cite{Raghu2015,Mahajan2013,Metlitski2015,Wang2016,Moon2010}, where the interaction is at its maximum, but doesn't
  distinguish between pairing channels. These studies found that $T_c$ is comparable to the upper boundary of the NFL behavior. The weak coupling FL study focused on the angular variation of the gap along the whole FS and on the difference between the pairing strength in different spin-singlet pairing channels. This study found that at a finite distance from a nematic transition (measured by the inverse correlation length $\xi^{-1}$ of  nematic fluctuations) $s-$ wave pairing wins over $d-$ wave and higher symmetry channels,  but the splitting between the coupling strength in different channels scales as $\xi^{-1}$ and vanishes at a QCP.
That work also found that, at a finite $\xi^{-1}$, there are two scales in the problem: the relevant momentum transfer in the gap equation is of order $\xi^{-1}$, but the gap varies at a larger scale $\xi^{-1/3}$. In the FL description, both scales collapse when $\xi$ diverges.

Our work unifies the strong coupling and weak coupling approaches. We analyze the pairing near a $Q= 0$ nematic QCP including both  the angular dependence of the nematic form-factor along the FS and the dynamics of the pairing interaction and associated self-energy $\Sigma (\theta, \omega_m)$.
We obtain $T_c$ in different pairing channels and the angular variation of the pairing gap by solving the linearized Eliashberg gap equation right at a QCP, where $\xi^{-1} =0$.
 We argue that the gap variation along the FS and the difference between the couplings in $s-$wave and $d-$wave channels are governed by a single dimensionless parameter $\lambda$, which is the ratio of the effective boson-fermion coupling and the Fermi energy, which we assume to be of order bandwidth.
At a metallic QCP, interaction is assumed to be smaller than the bandwidth, and we treat $\lambda$ as a small parameter.

We show that  $T_c$ remains finite at a QCP, and  $s-$wave and $d-$wave channels remain non-degenerate. The difference between the two comes from the dynamical part of the pairing interaction. The $T_c$ for $s-$wave pairing is higher, and the difference $1 - T^d_c/T^s_c \propto \lambda^2$.  We show that the angular dependence of the form-factor causes
 a sharp
angular variation of the pairing gap along the FS in both $s-$ and $d-$channels
 as a function of distance $\theta$ along the FS from where the form factor is maximal (i.e., from $ \theta = n \pi/2$). The pairing gap  is the largest in ``hot'' regions  with a width
of order  $\theta_{h} \sim \lambda^{1/3}$.  This scale is parametrically larger than the typical momentum transfer by the interaction, $O(\lambda)$, but smaller than typical scale of variation of the form-factor, which is $\theta = O(1)$. Between the two scales the gap behaves as $(\theta_{h}/\theta)^4$.
This behavior holds for both $s-$ wave and $d-$ wave pairing gaps, and the difference between the two develops
at $\theta = O(1)$.

\paragraph*{{\bf The Model.}}
\label{sec:model-results}

We base our study on the standard boson-fermion coupling model \cite{Hertz1976,Millis1993,Altshuler1994}. The bosons represent some collective degree of freedom, either charge excitations near a Pomeranchuk instability,  or some composite spin fluctuations responsible for d-wave nematic order. We assume a circular FS and dispersion $\epsilon_{\k} = k^2/2m -\mu$, but a generalization to a more general FS is straightforward.
The $d-$wave symmetry of a nematic order  is encoded in the fermion-boson interaction,
\begin{equation}
  \label{eq:H-int}
  H_I = g \sum_{\q,\k,\sigma}f(\k) \phi(\q)\psi_\sigma^\dagger\left(\k + \frac{\q}{2}\right)\psi_\sigma\left(\k - \frac{\q}{2}\right),
\end{equation}
in which $f(\k)$ represents the $d-$wave form-factor
and $\phi (q)$ is a bosonic field with static propagator $\chi (q) = \chi_0/(q^2 + \xi^{-2})$. At a QCP, $\xi^{-2} =0$.  The effective boson-fermion interaction is $\gb = g^2 \chi_0$ and the dimensionless coupling $\lambda \sim {\bar g}/E_F$.
In our problem, the relevant degrees of freedom are near the FS, so we approximate $f({\bf k})$ by an angular function $f(\theta_k) = \cos2\theta_k$.

We use as an input the result of earlier studies~\cite{Abanov2003,Oganesyan2001,Metlitski2010a,Maslov2010,Klein2018}
 that to leading order in $\lambda$ fermionic and bosonic self-energies  are given by one-loop expressions with free-fermion propagators. The bosonic self-energy gives rise to Landau damping and changes the bosonic propagator at a QCP to
\begin{equation}
  \label{eq:D-def}
  \chi(q,\theta_q,\W_m)^{-1} \approx \chi_0^{-1}\left(q^2 + \gamma f^2\left(\theta_q
    \right)\frac{|\W_m|}{\vf q}\right),
\end{equation}
where $\gamma = \gb m/\pi$ and $\gb = \chi_0 g^2$ is the effective coupling.
For fermions at the FS, the momentum transfer is $q = 2k_F \sin{\theta_q/2}$, and the susceptibility becomes the function of only $\theta_q$ and $\W$. The fermionic self-energy near the FS is
\begin{equation}
  \label{eq:se-def}
  \Sg(\theta_k,\w_m) = \w_0^{1/3}|f(\theta_k)|^{4/3}|\w_m|^{2/3}\sgn\w_m
\end{equation}
where
$\w_0 = (\gb/2\pi\sqrt{3})^3/\gamma\vf^2 \sim \gb^2/\ef
 \sim \lambda^2 \ef$.
The $\w^{2/3}$ form is a result of the $z = 3$ scaling.
\begin{figure}
  \centering
  \includegraphics[width=\hsize]{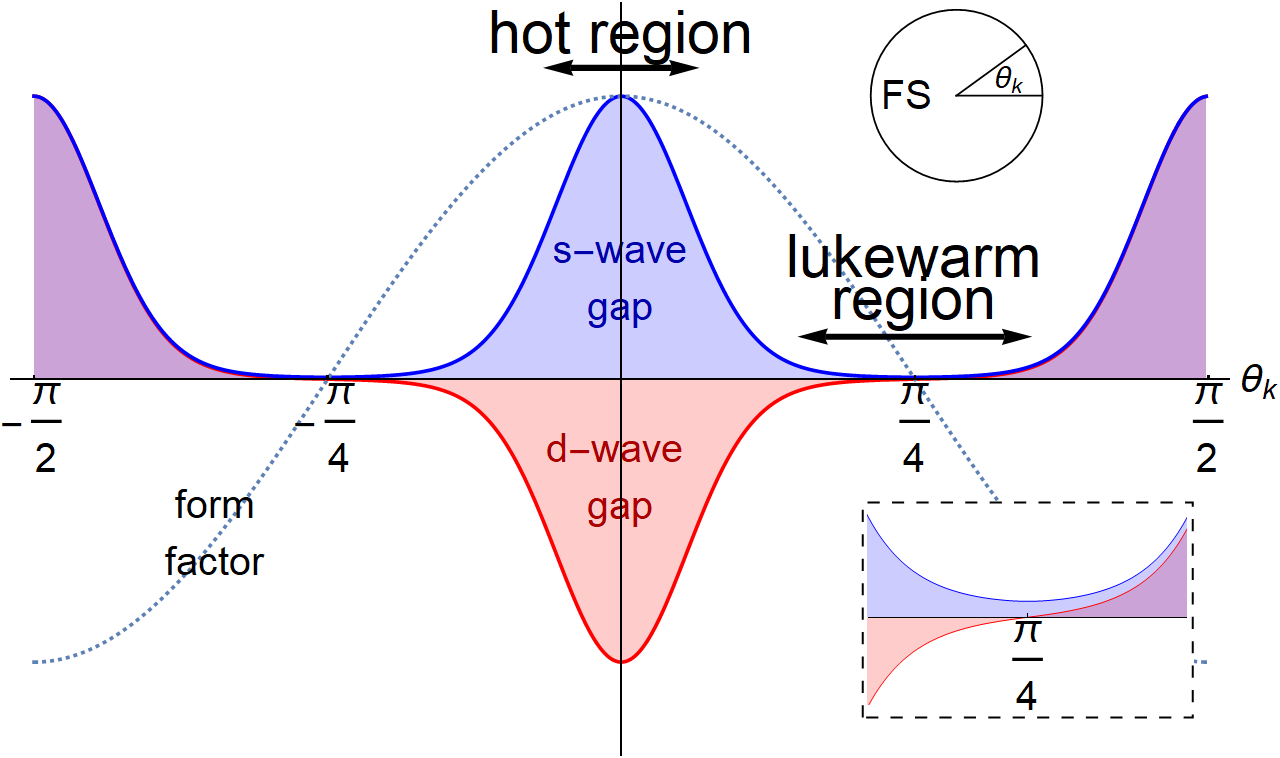}
  \caption{(color online) Behavior of the gap around the Fermi surface. The image depicts the numerical solution of the linearized Eliashberg gap equation \eqref{eq:gap-eq-1} at a nematic QCP, with the interaction form-factor $f(\theta_k) = \cos2\theta_k$ (dashed line). The blue (dark) and red (light) filled curves depict $s-$ wave and $d-$ wave solutions of the gap equation for weak coupling $\lambda = 0.025$. In both cases, the gap function is maximized in ``hot'' regions near $\theta = n \pi/2$, where the attraction is maximal. The width of a hot region is of order $\lambda^{1/3}$.
    Outside, the gap function rapidly drops and becomes of order $\lambda^{4/3}$,
    creating a ``lukewarm'' region (inset).
  }
  \label{fig:gaps}
\end{figure}

\paragraph{{\bf The Eliashberg equation.}}~~~
In order to obtain the linearized Eliashberg equation for the anomalous pair function $F(\theta_k,\w_n)$ we consider the ladder series of diagrams for infinitesimally small  $F(\theta_k,\w_n)$  with $g^2 \chi (q, \theta_q, \Omega_m)$ as the interaction and use full fermionic propagators with the self-energy $\Sigma (\theta_k, \omega_m)$. The Eliashberg equation is obtained by approximating the pairing interaction by that for fermions right on the FS (i.e., approximating $\chi (q, \theta_q, \Omega_m)$ by $\chi (\theta_q, \Omega_m)$  and
integrating out the momentum transverse to the FS in the fermionic propagators.
This is justified because typical bosonic momenta $q \sim \w^{1/3}$ are parametrically larger
than typical fermionic momenta $|k-k_F|
\sim \Sigma/v_F \sim \w^{2/3}$ for $\w  < \omega_0$ and $|k-k_F|
\sim \w/v_F$ for $\w > \omega_0$. Integrating over the  momentum transverse to the FS
we obtain
\begin{align}
  \label{eq:gap-eq-1}
  & F(\theta_k,\w_n) = \lambda
    T \sum_{\w_m\neq\w_n}\int_{-\pi}^{\pi}\frac{d\theta_q}{2\pi}\frac{F(\theta_k+\theta_q,\w_m)}{|\w_m + \Sg(\theta_k+\theta_q, \w_m)|} \times \nn \\
  &\qquad\qquad\frac{|2 \sin{\theta_q/2}| f^2\left(\theta_k + \theta_q/2\right)}{|2 \sin{\theta_q/2}|^3 + \frac{\gamma|\w_n-\w_m|}{\kf^3\vf}f^2\left(\theta_k + \theta_q/2\right)}
\end{align}
where we defined explicitly
\begin{equation}
  \label{eq:lambda-def}
  \lambda = \frac{\gb m}{2\kf^2}
  = \frac{\gb}{4\ef},~~~ \ef = \frac{\kf \vf}{2}.
\end{equation}
Notice that this  is a 2D integral equation in both frequency and the angle along the FS.  We removed the thermal contribution $\w_n = \w_m$, as it does not affect $T_c$ for spin-singlet pairing~\cite{Abanov2001,Abanov2008,Millis1988}, similar to the  effect  to non-magnetic impurities~\cite{Abrikosov1959,Abrikosov1975}.
Note that because $\Sigma (\theta_k, \omega_m) \propto \omega_m (\omega_0/\omega_m)^{1/3}$ and
$\gamma|\w_n-\w_m|/(\kf^3\vf) \propto \lambda^3 |\w_n-\w_m|/\omega_0$, Eq. (\ref{eq:gap-eq-1}) depends on a single parameter $\lambda$, when $T$  is rescaled by $\omega_0$.

Eq. \eqref{eq:gap-eq-1} has a straightforward interpretation. The $F/|\w + \Sg|$ term is the result of integrating out the fermionic particle-particle bubble, that for a constant interaction would give the usual $F/|\w_m|$ BCS form of the gap equation. The term on the second line is the bosonic susceptibility, weighted by the vertex form-factors, and $ 2k_F\sin(\theta_q/2)$ is momentum variation between two points on the FS separated by an angle $\theta_q$. For small angles, $ 2\sin(\theta_q/2)  \approx \theta_q$. Because of $f^2-$ factor in various places in the Eliashberg equation,
the FS can be
segmented into `hot' regions, where  $f^2(\theta) \simeq 1$, and `lukewarm' 
 regions where
$f^2(\theta) \ll 1$. Fig. \ref{fig:gaps} depicts the behavior of the form-factor and shows the hot and 
lukewarm
regions of the FS.

\paragraph{{\bf $T_c$ and the angular variation of $F(\theta_k, \w_m)$.}}~~~
We first obtain $T_c$. The frequency sum over $\omega_m$ in  (\ref{eq:gap-eq-1}) is UV
convergent, hence typical $\omega_n$ and $\omega_m$ are of the same order of $T_c$  Typical $\theta_q$ are then of order  $(\gamma|\w_n -\w_m|)^{1/3} \sim \lambda (T_c/\omega_0)^{1/3} $. We will see that in our case $T_c \sim \omega_0$. Then typical $\theta_q$ are of order $\lambda \ll 1$.   The $d-$wave form-factor does not vary on such scale and can be set to $f =1$.  We assume and then verify that $F (\theta_q + \theta_k, \omega_m)$ also varies slowly at $\theta_q = O(\lambda)$ and can be approximated by $F(\theta_k,\omega_m)$. In this situation we can integrate over $\theta_q$ in (\ref{eq:gap-eq-1}) and obtain a local gap equation,
\begin{equation}
  \label{eq:gap-local-2}
  F(\theta_k,n) \approx
  \sum_{m \neq n}
  F(\theta_k,m) \Lambda(m,n),
\end{equation}
where
\begin{equation}
  \label{eq:lambda-theta}
  \Lambda (m,n) =
  \frac{1}{3}
  \frac{1}{\left|m+\frac{1}{2}\right|^{2/3}|m - n|^{1/3}}\frac{1}{1+\left|2\pi T(m+\frac{1}{2})/\omega_0\right|^{1/3}}
\end{equation}
Eq. \eqref{eq:gap-local-2} is dimensionless, local, and universal in the sense that dimensionless $\lambda$ cancels out.
Solving Eq. (\ref{eq:gap-local-2}) numerically, we find
\begin{equation}
  2\pi T_c  = 2.9 \omega_0
  = 3.5 \times 10^{-3}
  \frac{\gb^2}{E_F}.
  \label{eq:ex_1}
\end{equation}
This is consistent \cite{strongSC2018_foot1} with earlier works~\cite{Bonesteel1996,Moon2010,Wang2016,Lee2018,Wu2018a}.

We next look at a
  lukewarm
   region and examine whether the interaction within this region can give rise to a comparable $T_c$. For definiteness let's focus on $\theta_k$ near $\pi/4$.
In lukewarm regions we need to differentiate between $s$-wave and $d-$wave (even and odd) solutions with
  $F^s(\theta_k,\w_n)  \approx F^s(\pi/4, \w_n), F^d(\theta_k,\w_n) \approx  F^d(\w_n)
  \delta \theta_k$,
  where $\delta \theta_k = \theta_k-\pi/4$.
  Because $f^2(\pi/4+\theta_q/2) = \sin^2{\theta_q}/2$, the effective static boson-mediated interaction $f^2 (\pi/4 + \theta_q/2) \chi (\pi/4 + \theta_q/2) = {\bar g}/(4k^2_F) = \lambda(2m)$ is not singular and weak.  In this situation, one can
  neglect both the Landau damping and the fermionic self-energy. Then $F^s(\pi/4,\w_n)$ does not depend on $\w_n$, i.e., the pairing is described by BCS theory,
  with an onset temperature
  $T^{luke}_{s} \propto e^{-1/\lambda_s}$
  , where $\lambda_s
  = O(\lambda)$.
  The temperature $T_s^{luke}$ is indeed much smaller than $T_c$ in Eq. (\ref{eq:ex_1}),
  and the same holds for d-wave pairing.
  This implies that s-wave SC in a lukward region is induced by that in the hot regions.

We now  determine the angular variation of the gap
in the hot regions.
For definiteness consider the segment $0 \leq 0 \leq \pi/4$. We label a characteristic $\theta$ at which $F(\theta_k, \omega_n)$ varies as $\theta_{h}$. At a first glance, $\theta_{h}$ should be of order one because $f(\theta)$ varies at $\theta = O(1)$.   However, we show that $
\theta_{h}$ is actually  parametrically smaller and is of order $\lambda^{1/3}$. To see this, we assume that $\theta_{h} \ll 1$ and then verify it. Because typical $\omega_m$ and $\omega_n$ in the Eliashberg equation are of order $T_c$, i.e., $\omega_n \sim T_c$ and $\gamma|\omega_m-\omega_n|/k^3_F v_F \sim \lambda^3$, we can
reduce the 2D integral equation (\ref{eq:gap-eq-1}) to a 1D equation on $\theta_k$:
\begin{equation}
  F(\theta_k) =  \frac{3 \sqrt{3}\lambda}{4} \int \frac{d \theta_q}{\pi} \frac{F(\theta_k + \theta_q) |\theta_q|}{|\theta_q|^3 + \lambda^3}
  f^2\left(\theta_k+\frac{\theta_q}{2}\right).
  \label{eq:ex_2}
\end{equation}
If we approximate $f^2(\theta_k+\theta_q/2)$ by $1$ and $ F(\theta_k)$ and $F(\theta_k + \theta_q)$ by $F(0)$, we see that  Eq.  (\ref{eq:ex_2})  reduces to an identity, as  should be for $T=T_c$. Going beyond this approximation, we expand $f^2(\theta_k+\theta_q/2)$ in (\ref{eq:ex_2}) as $1 - (\theta_k+\theta_q/2)^2/2$.For $\theta_k < \theta_{h}$ the second term in $f^2$ is irrelevant by construction, but for  $\theta_{h} \leq \theta_k \ll 1$ it plays a major role. Indeed, for these $\theta_k$ there are two contributions to the r.h.s. of (\ref{eq:ex_2}).  One comes from the integration over a narrow range $\theta_q \sim \lambda$ and yields $F(\theta_k) (1 - O(\theta^2_k))$. The other comes from the coupling to hot region, where
$F(\theta_k + \theta_q) \approx F(0)$.  Typical $\theta_q$ for this second contribution are  $\theta_q \sim -\theta_k$, i.e., they are parametrically larger than $\lambda$.
This second contribution is then of order $ \lambda F(0) \theta_{h}/\theta^2_k$.  Substituting the sum of the two contributions into the r.h.s. of (\ref{eq:ex_2}) we obtain
\begin{equation}
  F(\theta_k) \sim  F(0)  \lambda \frac{\theta_{h}}{\theta^4_k}
  \label{eq:ex_3}
\end{equation}
By construction, $F(\theta_k)$ is supposed to vary at $\theta_k \sim \theta_{h}$. This yields $\lambda \theta_{h} \sim \theta^4_{h}$, i.e.,
\begin{equation}
 \theta_{h} \sim \lambda^{1/3}.
 \label{eq:ex_4}
\end{equation}
This scale is in between the ``width'' of the interaction $\lambda$ and $\theta = O(1)$, at which $f(\theta)$ evolves.
We see from (\ref{eq:ex_3}) that at $\theta_{h} \leq \theta_k \ll 1$, $ F(\theta_k) \sim  F(0) (\theta_{h}/\theta_k)^4$. At $\theta_k = O(1)$ (in the 
lukewarm
region) $F(\theta_k) \sim F(0) \theta^4_{h} \sim F(0) \lambda^{4/3} \ll F(0)$. The behavior of  $F(\theta_k)$ in this region is different for $s-$wave and $d-$wave pairing (see below).
\begin{figure}
  \centering
  \includegraphics[width=\hsize]{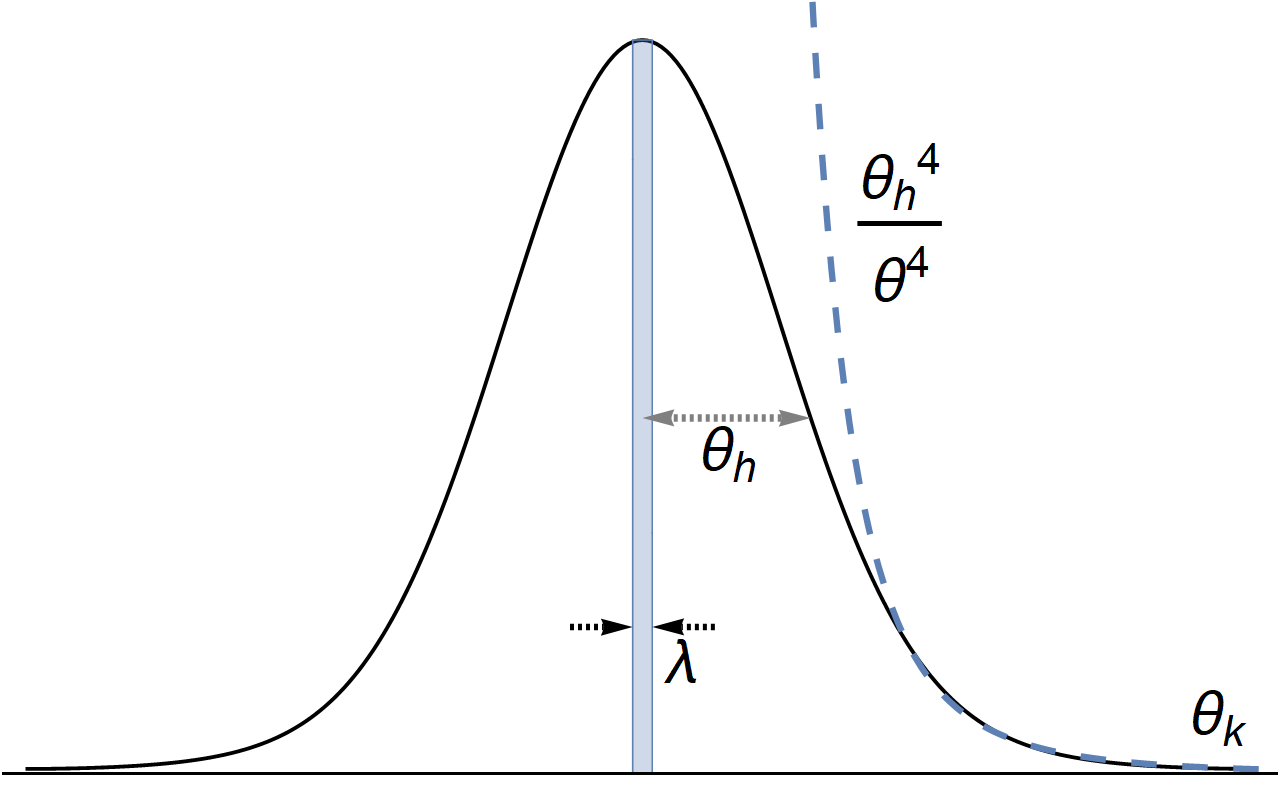}\llap{\makebox[\hsize][l]{\raisebox{0.35\hsize}{\includegraphics[width=0.42\hsize]{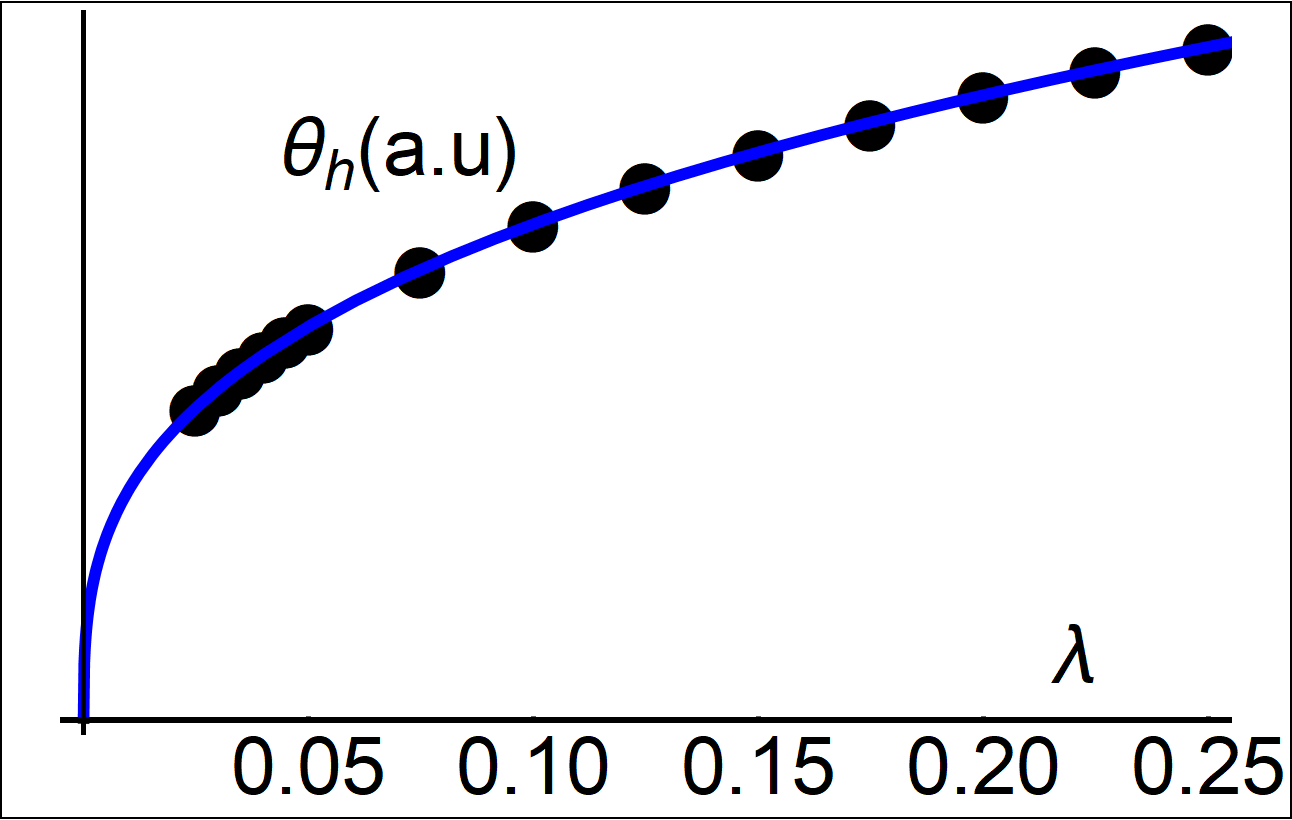}}}}
  \caption{Numerical solution of the full Eliashberg equation, Eq. \eqref{eq:gap-eq-1}, at small deviations from $\theta_p = 0$. Main panel  -- the gap function for $\lambda=0.025$. We define $2\theta_{h}$ as the full width at half-maximum. Insert -- the dependence of $\theta_{h}$ on $\lambda$. The solid line is a fit to $\lambda^{1/3}$. At $\theta > \theta_{h}$, the gap function scales as $(\theta_{h}/\theta)^{4}$, in agreement with Eq. \eqref{eq:ex_3}.}
  \label{fig:angular}
\end{figure}
In Fig. \ref{fig:angular}  we show the result of the numerical solution of the full 2D Eliashberg equation (\ref{eq:gap-eq-1}).
We see that for the full dynamical problem both the width of the interaction, and the width of the gap, are finite at a QCP. This is in contrast to a FL analysis \cite{Lederer2015}, where both vanish as $\xi^{-1},\xi^{-1/3}$ respectively, at a QCP.

\paragraph{{\bf $s-$wave vs $d-$wave pairing symmetry}}~~~
To obtain the global structure of the gap function and determine the splitting
 of onset temperatures $T_c^s, T_c^d$
for $s-$ wave vs $d-$ wave
 pairing, we need to take into account variations of the gap function over large regions of the FS,
 $|\theta_q| \sim \pi/2$.
Naively, we expect $s-$ and $d-$ wave splitting to be determined by whether the nematic attraction prefers a nodal
 $d$-wave structure or non-nodal $s-$wave structure.  However, it turns out that
 the condensation energy from $d-$wave nodes is 
 of order $\lambda^{11/3}$,
 and
 is much smaller than
 the actual $s-d$ energy difference
 which is of order $\lambda^2$.
 Instead, the splitting
     originates from the difference in the interactions   
  between 
  hot regions. To 
    show  this,
we again reduce the 2D integral equation
(\ref{eq:gap-eq-1}) to the effective 1D equation on $\theta_k$, as in Eq. \eqref{eq:ex_2},
but now do not expand the r.h.s. in small $\theta_k$ and $\theta_q$. The full effective 1D equation differs from
\eqref{eq:ex_2}, and this difference can be modeled by introducing eigenvalues $\eta_{s,d} \neq 1$, different for $s-$wave and $d-$wave pairing. Setting $\theta_k =0$, we then obtain
\begin{equation}
  \label{eq:split-1}
  \eta_{s,d} F(0) =
  \frac{3 \sqrt{3}\lambda}{4} \int \frac{d \theta_q}{\pi} \frac{F(\theta_q) |2\sin\theta_q/2|}{|2\sin\theta_q/2|^3 + \lambda^3}  f^2\left(\frac{\theta_q}{2}\right).
\end{equation}
One can verify that larger eigenvalue corresponds to larger $T_c$. Our goal
 is to find
  $\eta_s - \eta_d$.

The leading contribution to the r.h.s. of (\ref{eq:split-1}) comes from $\theta_q \leq \lambda$. This leading term, however, does not differentiate between $s-$wave and $d-$wave pairings. The one which differentiates between the two comes from the range of order $\theta_{h}$ near $|\theta_q| = \pi/2$.
This contribution is of order $\lambda\theta_{h}^3 \sim \lambda^2$ (the additional $\theta^2_{h}$ is due to $f^2(\theta_q/2) \propto \theta^2_{h}$  in the region $\theta_q \sim \pm \pi/2$). Accordingly, the splitting between $s-$wave and $d-$wave couplings is
\begin{equation}
  \label{eq:splitting}
  \eta_s - \eta_d \sim \lambda^2\sim \frac{\w_0}{\ef} \sim \frac{T_c}{\ef}.
\end{equation}
The eigenvalue splitting  gives rise to the splitting between $T^s_c$ and $T^d_c$: $(T^s_c -T^d_c)/T^s_c \sim \eta_s - \eta_d \propto \lambda^2$
(i.e., $T^s_c -T^d_c \propto E_F \lambda^4$).
One can verify that the higher eigenvalue is $\eta_s$,
as one expects considering that the interaction is purely attractive.
We also verified Eq. \eqref{eq:splitting} by numerically solving Eq. \eqref{eq:gap-eq-1} \cite{nematicSupp2018}.

\begin{figure}[t!]
  \centering
  \includegraphics[width=0.7\hsize]{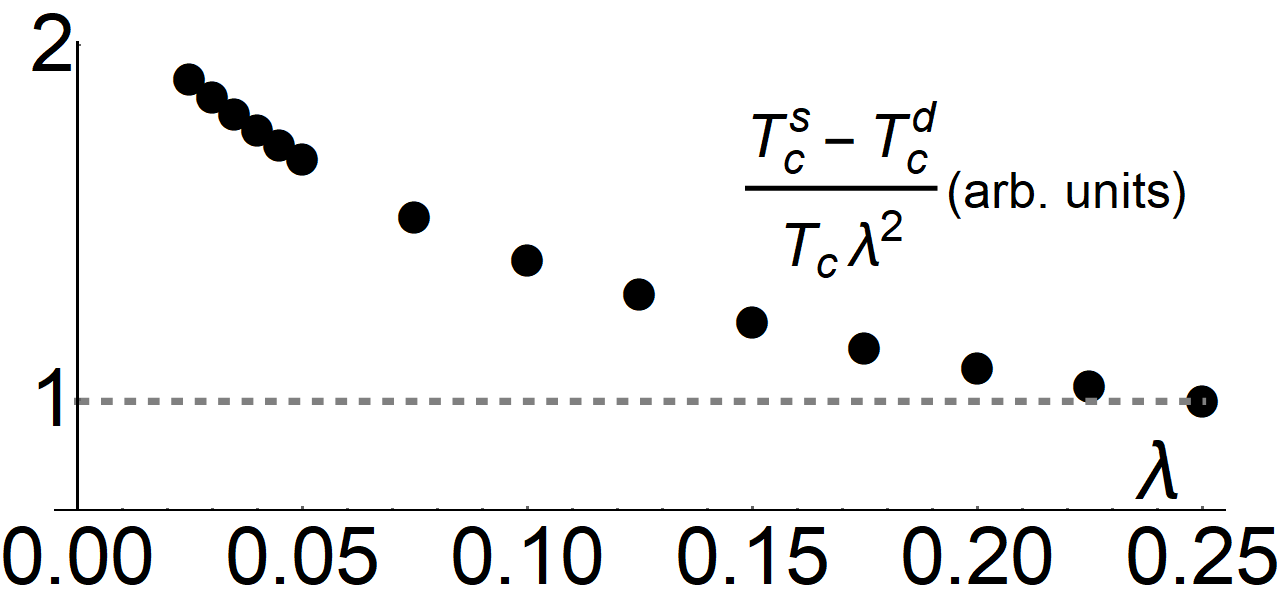}
  \caption{The
   splitting of $T^s_c$  and $T^d_c$ as a function of $\lambda$ from the solution of the full 2D Eliashberg equation. We plot the ratio
     $(T^s_c- T^d_c)/(T^s_c \lambda^2)$, normalized to 1 at $\lambda =0.25$.
  The result agrees with  Eq. \eqref{eq:splitting}.}
\end{figure}
Eqs. \eqref{eq:ex_1} and \eqref{eq:splitting} portray the interplay between long- and short- scales near a QCP. The divergence of static fluctuations near the QCP is cut off by the boson dynamics, setting the IR scale of momentum transfer $\theta_q \sim \lambda$. Interactions at this scale provide the largest contribution, of order $\omega_0 \sim \lambda^2 E_F$, to $T_c$ in both $s-$wave and $d-$wave channels. The degeneracy between $T_c$ in the two channels is lifted by the much weaker interaction at large momentum transfer of $\theta_q \sim 1$, and has additional smallness in $\lambda^2$.
  Note that although $\eta_s - \eta_d \propto \lambda^2$, the  
  the effective 1D equation \eqref{eq:split-1}, from which we extracted  $\eta_s - \eta_d$,  neglects self-energies at large angles, so we don't need to compute
  self-energies to order $\lambda^2$.

\paragraph{{\bf Summary.}}~~~
In this communication we studied  strong coupling theory of SC in a metal near a QCP towards $q=0$ nematic order. We used fermion-boson model, and treated the ratio of effective boson-fermion coupling and the Fermi energy as a small parameter $\lambda$. We solved the linearized Eliashberg equation and verified that $T_c$ is finite at a QCP and is of order $\lambda^2 E_F$ for both $s-$wave and $d-$wave pairing. The two are not degenerate and $T^s_c$ is larger than $T^d_c$, but the difference $T^s_c-T^d_c \sim \lambda^4 E_F$ is much smaller than each of these temperatures.  We also analyzed angular variation of the superconducting order parameter $F(\theta_k)$ along the FS. We showed that $F(\theta_k)$ is the largest
  in hot regions
 on the FS, whose width $\theta_{h} \sim \lambda^{1/3}$.
Within
 a hot region (at $\theta_k < \theta_{h}$),
the order parameter is approximately a constant. Outside, it drops as $(\theta_{h}/\theta_k)^4$ and becomes smaller by a factor $\lambda^{4/3}$.  This behavior holds for both $s-$wave and $d-$wave order parameters. The two become different only at
 $\theta_k = O(1)$.

We end with
a word of caution. In this work we considered $F(\theta_k)$ which monotonically decreases between hot and lukewarm regions and does not change sign along the arc $0< \theta_k < \pi/4$.
There exist other $s-$wave and $d-$wave solutions of Eq. (\ref{eq:gap-eq-1}), which change
 sign $n \geq 1$ times. These additional solutions emerge at smaller $T$ and do not affect $T^s_c$, $T^d_c$, and the structure of $F(\theta_k)$ near $T_c$ in each channel. Still, if $T_c$ for these additional solutions of the linearized equation is  small compared to $T_c$ only by some power of $\lambda$,
 we expect that
the form of $F(\theta_k)$ near $T=0$ will be quite different from that near $T_c$.
\begin{acknowledgments}
  We thank E. Berg, R. Fernandes, S. Kivelson, M. N. Gastiasoro, S. Lederer and Y. Schattner for stimulating discussions. This work was supported  by the NSF DMR-1523036.
  We acknowledge the Minnesota Supercomputing Institute at the University of Minnesota for providing resources that assisted with this work.
\end{acknowledgments}

\bibliography{QCP,NZS,nematicSC}
\clearpage
\onecolumngrid

  \section*{Supplementary material}
  \label{sec:suppl-mater}

  Our supplemenary material has two parts. The first part gives a more detailed derivation of our results on angular variation of the gap function $F(\theta_k)$ in both hot and lukewarm regions, and on the resulting splitting of critical temperatures $T^{s,d}_c$ between $s-$ wave $d-$ wave modes, Eq. \eqref{eq:splitting}. The second part discusses the numerical methods used to determine the critical temperature at the QCP, Eq. \eqref{eq:ex_1}, and to verify our analytic results.

  \subsection*{Angular variation of $F(\theta_k)$}
  \label{sec:behavior-gap-at}

  In the main part of the paper, we noted that the critical temperature is, to first approximation, determined by the local, frequency dependent, gap equation \eqref{eq:gap-local-2}. In order to determine the angular behavior, we approximated the full gap equation \eqref{eq:gap-eq-1} by an effective one dimensional integral equation where we replaced the frequency terms in the gap equation by their typical value $\w_n,\w_m \sim T_c$, and summed over the Matsubara frequencies. The result is Eq. \eqref{eq:ex_2} which we reproduce here for clarity,
  \begin{equation}
    F(\theta_k) =  \frac{3 \sqrt{3}\lambda}{4} \int \frac{d \theta_q}{\pi} \frac{F(\theta_k + \theta_q) |\theta_q|}{|\theta_q|^3 + \lambda^3}
    f^2\left(\theta_k+\frac{\theta_q}{2}\right).
    \label{eq:supp-1}
  \end{equation}
  Eq. \eqref{eq:supp-1} neglects several angular terms, namely the angular dependency of the fermionic and bosonic self-energies, see Eqs. \eqref{eq:D-def}, \eqref{eq:se-def}. We have verified that neglecting these terms doesn't affect the final result. Eq. \eqref{eq:supp-1} has been the property that if we neglect the dependence of $F$ and $f^2$ on $\theta_q$, it is fulfilled trivially.

  To determine the width of the hot region gap we assume that $F = F(\theta_k/\theta_{h})$ is a function of a single scaling parameter $\theta_{h}$, and analyze it for $1 \gg \theta_k \gg \theta_{h}$. The r.h.s. of Eq. \eqref{eq:supp-1} simplifies to,
  \begin{align}
    \label{eq:supp-2}
    0 &\approx -F(x)\theta_{h}^2x^2/2 + \frac{3 \sqrt{3}\lambda}{4\pi \theta_{h}} \int dy  \frac{F(y)}{|x-y|^2},
  \end{align}
  where $x = \theta_k/\theta_{h} \gg 1$, but $\theta_{h}^2x^2 \ll 1$. The first term is the local contribution from $\theta_q \sim \lambda$, and the second term is the induced gap from the nearby hot region at $\theta_q \sim -\theta_k$.  It is easy to see that for
  \begin{equation}
    \theta_{h}^3 = \frac{3\sqrt{3}\lambda}{2\pi}
  \end{equation}
  we obtain a dimensionless equation (for $x\gg 1$),
  \begin{equation}
    \label{eq:supp-approx-eq-hs}
    F(x)  = \frac{1}{x^2}\int dy \frac{F(y)}{(x-y)^2}
  \end{equation}
  with a solution,
  \begin{equation}
    \label{eq:supp-3}
    F(x) \approx a F(0)/x^4,
  \end{equation}
  where $a$ is a constant of order one.
  Our results are equivalent to Eqs. \eqref{eq:ex_3},\eqref{eq:ex_4}.
  Eq. \eqref{eq:supp-3} also demonstrates that near the lukewarm regions $\theta_k \sim 1$,
  \begin{equation}
    \label{eq:cold-amp}
    F(\theta_k) \sim F(0)\theta_{h}^4 \propto F(0) \lambda^{4/3}.
  \end{equation}

  In order to obtain the transition temperatures for $s-$ wave and $d-$ wave gaps, we again reduce Eq. \eqref{eq:gap-eq-1} to an effective 1D equation. We account for the expected temperature differences by introducing different eigenvalues for $s-$ wave and $d-$ wave solutions $\eta_s(T), \eta_d(T)$, i.e.,
  \begin{equation}
    \eta_{s,d}F(\theta_k)_{s,d} =  \frac{3 \sqrt{3}\lambda}{4} \int \frac{d \theta_q}{\pi} \frac{F_{s,d}(\theta_k + \theta_q) |2\sin\theta_q/2|}{|2\sin\theta_q/2|^3 + \lambda^3}
    f^2\left(\theta_k+\frac{\theta_q}{2}\right).
    \label{eq:supp-4}
  \end{equation}
  We assume and then verify that $(T_c^s-T_c^d)\ll T_c$, and expand the $\eta$'s near $T_c^s,T_c^d$, to obtain,
  \begin{equation}
    \label{eq:supp-eta-tc}
    \eta^{s,d}_c(T_c) \approx 1 + \alpha_{s,d}\frac{T^{s,d}_c-T_c}{T_c},
  \end{equation}
  where $T_c$ is the solution, Eq. \eqref{eq:ex_1}, of the local gap equation \eqref{eq:gap-local-2}. Then we have
  \begin{equation}
    \label{eq:supp-ts-td}
    1 - \frac{T_c^d}{T_c^s} \approx \frac{\eta_s-1}{\alpha_d} - \frac{\eta_d - 1}{\alpha_s}.
  \end{equation}
  In order to evaluate $\eta_{s,d}$ we again account for the two contributions from the r.h.s. of Eq. \eqref{eq:supp-3}, one coming from the local contribution $\theta_q \sim 0$, and the other coming from far regions, $|\theta_q| \gg \theta_{h}$. The local contribution is larger, but doesn't differentiate between $s-$wave and $d-$ wave, which will be determined by the nonlocal contribution. If we consider the behavior at a hot region, say $\theta_k = 0$, then the nonlocal contribution will come mostly from the hot regions at $\theta_q = \pm \pi/2$. Therefore we have,
  \begin{align}
    \label{supp-split-1}
    \eta_{s,d} F(\theta_k = 0) &\approx F(0) \pm 2\int \frac{3\sqrt{3}\lambda}{8\pi}\int d\theta_q  F(\theta_q)f^2\left(\frac{\pi}{4}+\frac{\theta_q}{2}\right) \nn\\
                               &\approx F(0) \pm a\lambda\theta_{h}^3F(0).
  \end{align}
  where in the integration we shifted $\theta_q \to \theta_q\pm \pi/2$. In the second line, one $\theta_{h}$ in the last term on the right comes from width of the hot region, and another $\theta_{h}^2$ comes from expanding the form-factor, $f^2(\pi/4+\theta_q/2) \approx \theta_q^2/4$. $a$ is a constant of order one. Eq. \eqref{supp-split-1} implies a splitting $\eta_s - \eta_d \sim \lambda^2$, which is second order in $\lambda$. Such splitting is much smaller than what we would naively expect, namely a difference of order $\lambda$. We therefore need to verify that there is no other contribution that is equivalent or larger. To this end we re-iterate Eq. \eqref{eq:supp-3}, and obtain for $\theta_k = 0$,
  \begin{align}
    \label{eq:supp-split-2}
    \lambda_{s,d}^2 F(0) &= \left(\frac{3 \sqrt{3}\lambda}{4}\right)^2 \int \frac{d \theta_q}{\pi}\frac{d \theta_q'}{\pi} \frac{F(\theta_q+\theta_q') |2\sin\theta_q'/2|}{|2\sin\theta_q'/2|^3 + \lambda^3}
        f^2\left(\frac{\theta_q+\theta_q'}{2}\right)\frac{|2\sin\theta_q/2|}{|2\sin\theta_q/2|^3 + \lambda^3}  f^2\left(\frac{\theta_q}{2}\right) \nn\\
                         &\approx F(0)(1 \pm 2a \lambda \theta_{h}^3 + b_\pm \lambda^2 \theta_{h})
  \end{align}
  Here $b_\pm$ are constants of order one. The final term comes from one of two contributions: (a) $\theta_q \sim 0$ but $0 \ll |\theta_q'| \ll \pi/2$, or vice versa. This is a contribution from the lukewarm region. (b) $0 \ll |\theta_q|,|\theta_q'| \ll \pi/2$, but $|\theta_q + \theta_q'| \sim 0,\pi/2$. This is a contribution from the hot regions. Regardless of origin, the final contribution is clearly smaller than the second term, and so, going back to Eq. \eqref{eq:supp-ts-td}, we find that the split in $T_c^s, T_c^d$ scales with $\lambda^2$. Eq. \eqref{eq:supp-ts-td} is equivalent to Eq. \eqref{eq:splitting} in the main text.

  \subsection*{Numerical methods}
  \label{sec:numerical-methods}

  We performed numerical analysis of the two gap equations we studied in the main text: both the full 2D Eliashberg equation, Eq. \eqref{eq:gap-eq-1}, and the local gap equation, Eq. \eqref{eq:gap-local-2}. All of our solutions were obtained in MATLAB 2017.

\begin{figure}
    \centering
    \includegraphics[width=0.8\hsize]{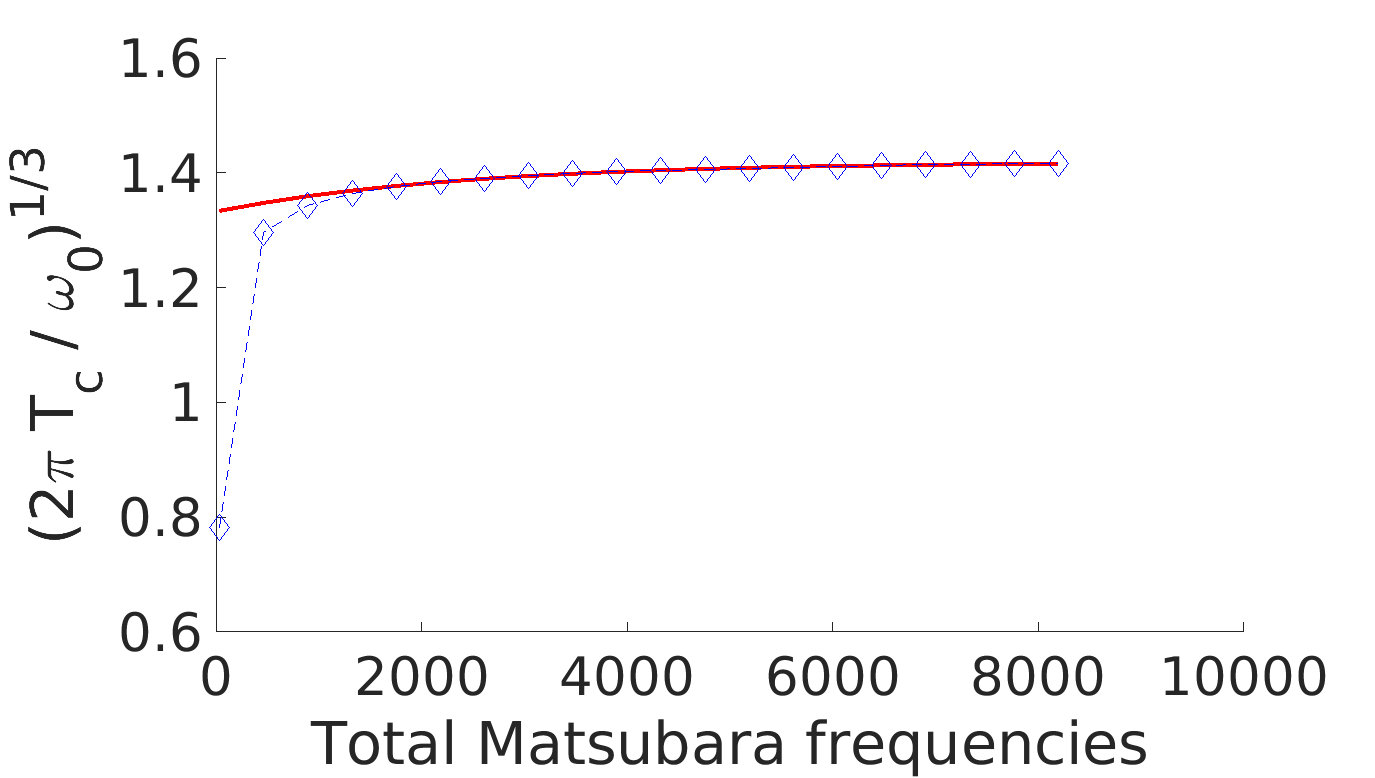}
    \caption{Scaling of $T_c$ in the local gap equation as a function of number of Matsubara frequencies included in the summation. The solid red line is a fit to $a + b \exp(-c x)$. The extrapolated result is reported in
      Eq. \eqref{eq:ex_1} of the main text.}
    \label{fig:tc-scaling-matsubara}
  \end{figure}

  We solved the local gap equation by numerically finding the largest eigenvalue of the operator on the r.h.s. of  Eq. \eqref{eq:gap-local-2}. We solved for using an increasing series of Matsubara frequencies, and then performed finite-size scaling. The result is shown in Fig. \ref{fig:tc-scaling-matsubara} and was reported in Eq. \eqref{eq:ex_1} of the main text.

  We solved the full 2D Eliashberg gap equation for a variety of of system sizes in both angle discretization and Matsubara frequencies, $N_\theta = 2^7-2^9$, $N_M = 2^3-2^6$, and a variety of couplings, $\lambda = 0.025-0.25$. All computations were performed using the resources of the Minnesota Supercomputing Institute (MSI). We confirmed numerically the calculated scaling of the hot region width and decay, Eqs. \eqref{eq:ex_3}, \eqref{eq:ex_4}. We also confirmed that the eigenvalue splitting between $s-$ wave and $d-$ wave solutions of the full equation followed the same scaling as the one we found from the 1D equation, Eq. \eqref{eq:splitting}. We also confirmed the expected height of the gap in the lukewarm region, Eq. \eqref{eq:cold-amp}.
\end{document}